\documentclass[summary]{ursi}

\usepackage{enumitem}
\usepackage[colorlinks=true, linktocpage, linkcolor={black}, citecolor={black}, urlcolor={black}]{hyperref}

\title{Progress in the Design of the Atacama Large Aperture Submillimeter Telescope}

\author{
Tony Mroczkowski\affref{ref1}, 
Claudia Cicone\affref{ref2},
Matthias Reichert\affref{ref6}, 
Patricio Gallardo\affref{ref4},
Hans Kaercher\affref{ref3}, 
Richard Hills\affref{ref5},
Daniel Bok\affref{ref6}, 
Erik Dahl\affref{ref6},
Pierre Dubois-dit-Bonclaude\affref{ref6},
Aleksej Kiselev\affref{ref6}, 
Martin Timpe\affref{ref6}, 
Thomas Zimmerer\affref{ref6},
Simon Dicker\affref{ref7},
Mike Macintosh\affref{ref8},
Pamela Klaassen\affref{ref8},
Michael Niemack\affref{ref9}\affref{ref10}
}

\affiliation{%
  \aff{ref1}{European Southern Observatory, Karl-Schwarzschild-Str.\ 2, Garching 85748,
Germany; e-mail: tonym@eso.org}
  \aff{ref2}{Institute of Theoretical Astrophysics, University of Oslo, P.O. Box 1029, Blindern, 0315 Oslo, Norway}
  \aff{ref6}{OHB Digital Connect, Weberstra\ss e 21, D-55130 Mainz, Germany}
  \aff{ref4}{Kavli Institute for Cosmological Physics, University of Chicago, Chicago, IL, 60637, USA}
  \aff{ref3}{Independent Consultant, Kirchgasse 4, D-61184 Karben, Germany}
  \aff{ref5}{University of Cambridge, Cambridge, UK; Deceased June 5, 2022.}
  \aff{ref7}{Department of Physics and Astronomy, University of Pennsylvania, 209 South 33rd Street, Philadelphia, PA, 19104, USA}
  \aff{ref8}{UK Astronomy Technology Centre, Royal Observatory Edinburgh, Blackford Hill, Edinburgh EH9 3HJ, UK}
  \aff{ref9}{Department of Physics, Cornell University, Ithaca, NY 14853, USA}
  \aff{ref10}{Department of Astronomy, Cornell University, Ithaca, NY 14853, USA}
}

\begin{document}

\maketitle

\begin{abstract}
  The Atacama Large Aperture Submillimeter Telescope (AtLAST) aims to be the premier next generation large diameter (50 meter) single dish observatory capable of observations across the millimeter/submillimeter spectrum, from 30~GHz to 1~THz.  
  AtLAST will be sited in Chile at approximately 5100 meters above sea level, high in the Atacama Desert near Llano de Chajnantor.  
  The novel rocking-chair telescope design allows for a unprecedentedly wide field of view (FoV) of 1-2$^\circ$ diameter, a large receiver cabin housing six major instruments, and high structural stability during fast scanning operations (up to $\sim 3^\circ$ per second in azimuth).
  Here we describe the current status of, and expected outcomes for, the antenna design study, which will be completed in 2024.
\end{abstract}

\section{Introduction}

Observations in the millimeter/submillimeter (mm/submm) probe astrophysical environments, such as the cold and dense molecular gas out of which stars and planets are formed and the (now extremely cold, $T=2.73$~K) Cosmic Microwave Background (CMB), which are otherwise invisible at wavelengths accessible to the human eye. Submm/mm observations are crucial to study the gas flows and energetic phenomena that shape the formation and evolution of galaxies across cosmic times as well as the origins of life. An overview of the some of these science goals can be found in \cite{2022SPIE12190E..07R} and references therein.

Despite the unique and transformational science needs for such a facility, no submm/mm observatories currently exist that have sufficiently large fields of view (FoVs) to map large portions of the sky while simultaneously providing sufficient angular and spectral resolution to resolve extragalactic and Galactic sources.  On the one extreme, the Atacama Large Millimeter/submillimeter Array (ALMA), delivers sensitive observations at up to milliarcsecond resolution in its most extended configurations, but due to the small FoV and spatial filtering intrinsic to interferometers, suffers poor sensitivity to extended diffuse emission and limited survey capabilities.
Upgrades to ALMA planned for this and next decade \cite{2022arXiv221100195C} could expand the bandwidth (2-4$\times$) and improve correlator, digitizer, and receiver sensitivities, resulting in $3-6\times$ continuum and $2-3\times$ for line improvements in imaging speed. 
However, applying these factors to current large ALMA program surveys still implies they would cover less than the solid angle subtended by the full Moon, and would still fail to recover extended scales.
On the other extreme, sub-10-meter class single dish telescopes with very large throughputs (instantaneous FoV times collecting area) like Simons Observatory, the Fred Young Submm Telescope (FYST) \cite{2018SPIE10700E..41P}, and CMB-S4 will soon survey most of the sky, but at low spatial resolution.  
A large aperture ($>$25-m), large FoV single dish  telescope is therefore crucially needed.

The Atacama Large Aperture Submillimeter Telescope (AtLAST)\footnote{\url{https://atlast-telescope.org/}}, providing a FoV of $1-2^\circ$ diameter, would be the largest submm/mm single dish in the Southern Hemisphere.
For comparison, the 50-meter Gran Telescopio Milim\'etrico (GTM) Alfonso Serrano,\footnote{\url{http://lmtgtm.org/}} operates at 0.85-4~mm (75-350~GHz), has FoV$\, \approx 0.07^\circ$ in diameter, and is located at $\approx 19^\circ$N latitude, on a site where the median precipitable water vapor rarely falls below 2~mm, limiting submm observations.

AtLAST therefore presents a powerful complement to ALMA and to upcoming cosmological survey telescopes and a much needed transformative next generational step. The AtLAST design study is funded by the EU\footnote{\url{https://cordis.europa.eu/project/id/951815}} and led by the University of Oslo, with participation by ESO, OHB Digital Connect, the UK Astronomy Technology Centre, the University of Hertfordshire, and support from a large international community (see \cite{2022SPIE12190E..07R}).

\section{Historical context}

Large single dish telescopes are not a new concept, especially at lower frequencies. Historically, however, as radio astronomers sought higher resolutions and more collecting area, the concept of an interferometer as a `broken mirror' that would provide high resolution without the cost and mechanical complexity of building gigantic structures was adopted.  In terms of resolution, no single dish can compete with e.g.\ ALMA or global mm-VLBI networks.  However, many of these interferometric approaches were envisioned at a time when widefield heterodyne \cite{2019arXiv190703479G} and large bolometer arrays did not exist (see for example the ALMA Memo Series \footnote{\url{https://library.nrao.edu/alma.shtml}}, which dates back to 1982). 
 
The AtLAST project builds on the experience from many large single dishes of the past, but it is highly innovative in that its large FoV will ultimately require gigapixel arrays and so enable the observatory to push technological boundaries for decades to come. Further, AtLAST will achieve a factor $>2.5\times$ higher frequency observations than any current $>15$-m mm/submm telescope. 
As discussed in \cite{2020SPIE11445E..2FK}, the state of the art for detector technologies advances rapidly, with detector counts increasing roughly $10\times$ every 7 years. Provided this scaling holds, the 2030s will see the first megapixel detector arrays, while over the operational lifetime of AtLAST we can expect to reach tens of gigapixels, perhaps more.  Meanwhile, sustainability considerations strongly motivate a design for AtLAST that can be maintained and upgraded for many decades.

\section{Design description}

\begin{figure}[htb]
  \centering
  \includegraphics[width=0.8\columnwidth]{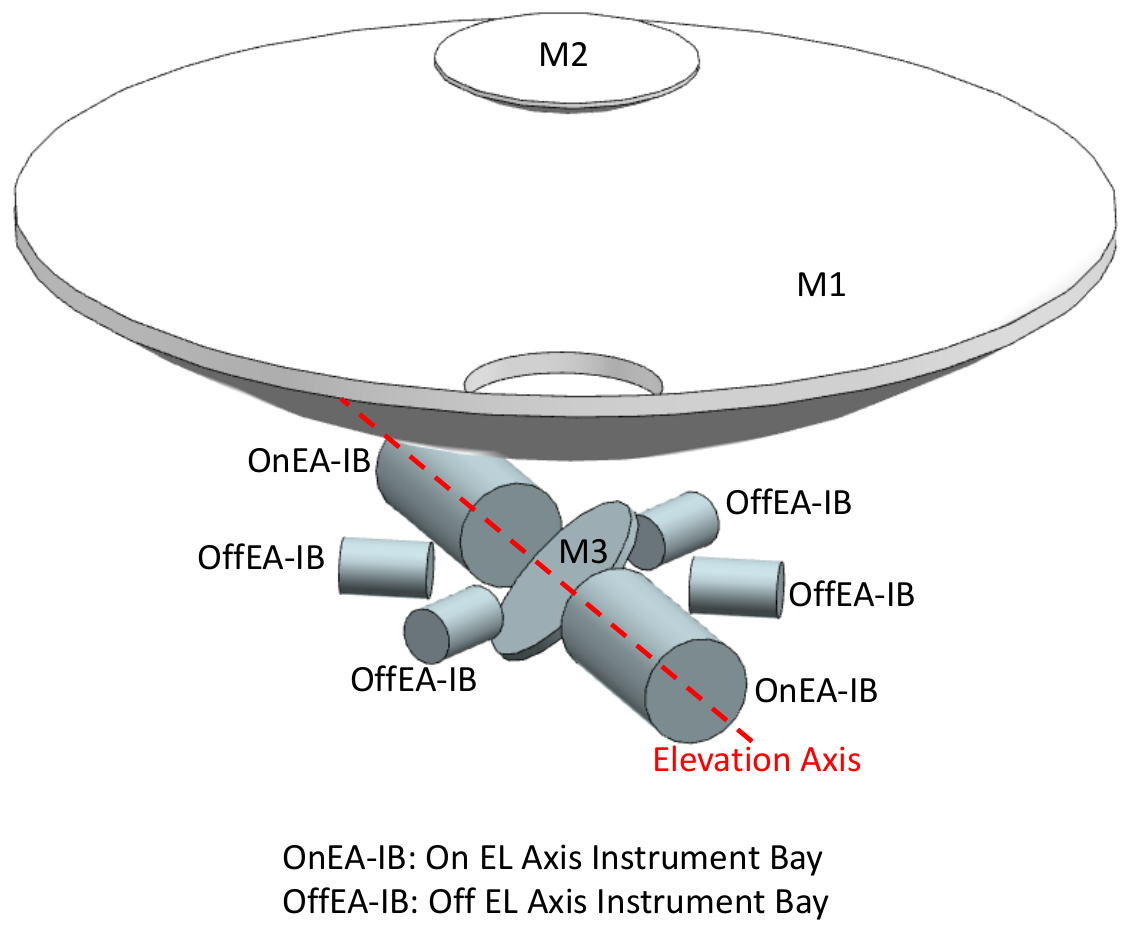}
  \caption{Diagram of the six major instrument locations in the AtLAST receiver cabin. }
  \label{fig:inst}
\end{figure}

\begin{figure*}[hbt]
  \centering
  \includegraphics[clip,trim=0mm 0mm 0mm 0mm,width=0.495\textwidth]{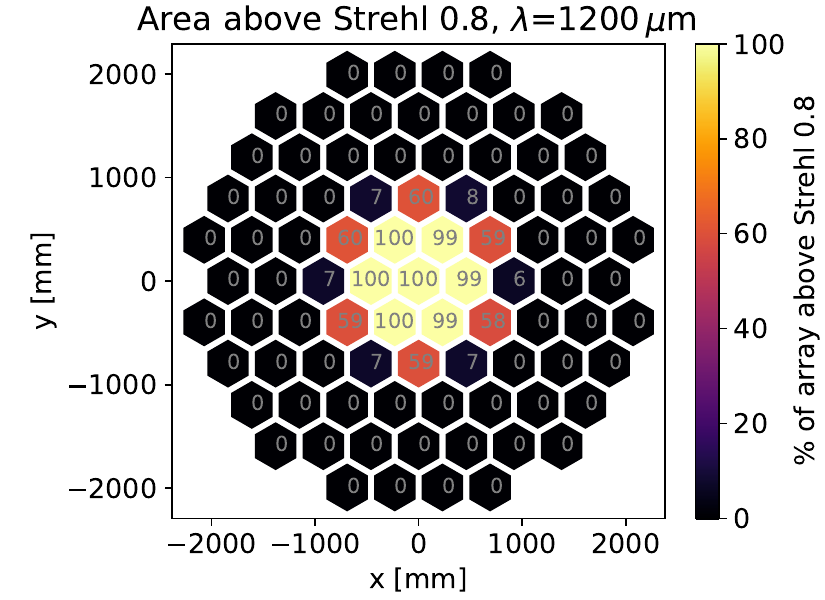}
  \includegraphics[clip,trim=0mm 0mm 0mm 0mm,width=0.495\textwidth]{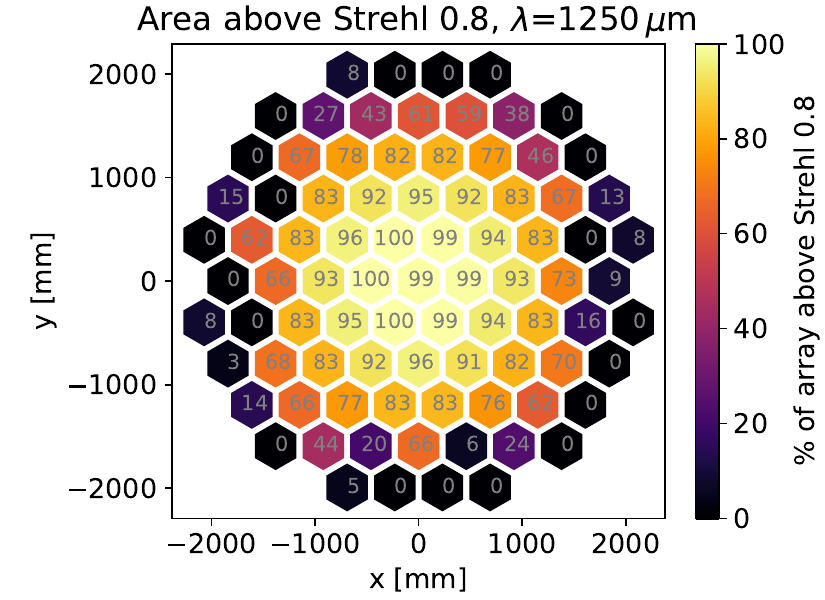}
  \caption{Area of the focal plane with Strehl ratio above 80\%, shown at a representative frequency of 1.2~mm. The left panel shows that, if uncorrected, only six out of 85 optics tubes could be used for observations.  The right panel, which results by adding biconic lenses similar to those used in CMB-S4 \cite{Gallardo2022}, shows a dramatic improvement whereby $>40\%$ of the focal plane becomes usable.}
  \label{fig:strehls}
\end{figure*}

The AtLAST design study covers a broad range of activities, including the project governance, site selection, operations models, design of an off-grid sustainable power system, and further development of the science case.  In this summary, we focus the telescope design itself, work that is primarily carried out at OHB DC. The study covers the structural, mechanical, and electrical design, full end-to-end simulations, solutions for a closed-loop active telescope optics, and will culminate in a preliminary design and construction plan for the telescope.

AtLAST takes an on-axis Cassegrain optics approach (see \cite{AtLAST_memo_1}) with a baseline design using a 50-m primary (M1) and 12-m secondary (M2), shown in Figure~\ref{fig:inst}.  A flat tertiary mirror (M3) allows selection of the instrument at the focus and for two of the six instruments to be in a Nasmyth-mounted configuration, bringing the advantage that they do not move in elevation.
The two Nasmyth-mounted instruments are located in instrument bays (IBs) along the elevation axis (EA), labeled OnEA-IB in Figure \ref{fig:inst}, and four smaller off-axis Cassegrain mounted instruments, labeled OffEA-IB.  
The OnEA instrument bays are designed to accommodate (up to 25 ton) instruments that use the entire 2$^\circ$ FoV, corresponding to a (curved) 4.3 meter diameter focal plane.
The 4 smaller OffEA instruments can each accommodate up to a 1$^\circ$ FoV, corresponding to instruments up to 2.1 meters in diameter.  We note that the overall throughput of each OffEA instrument will be comparable to that of a fully instrumented FYST, which has a 6-meter primary and 8$^\circ$ diameter FoV. 
While tipping in elevation can be non-ideal in terms of instrument design, we note a number of instruments have demonstrated cryogenics that achieve this, including for example the receivers on ALMA as well as both MUSTANG and MUSTANG-2 on the 100-meter Green Bank Telescope.

As discussed in AtLAST Memo \#1 (\cite{AtLAST_memo_1}), the primary motivation for the on-axis approach is to maximize the size of M1 while retaining a compact format that minimizes the overall mass and cross-section of the telescope. 
The fast focal ratio of $\approx 2.6$ F/\# and the large M2 allow geometrically for a large FoV.
The on-axis design choice presents challenges for realizing a large FoV, which we address by using an approach similar to that taken for instrumentation on the 6-m elements of CMB-S4 \cite{Gallardo2022}.
We note that the Strehl ratio, which quantifies the beam quality, remains acceptable ($>0.8$) over larger portions of the field for lower frequencies. 
The results of the Strehl ratio calculations with and without corrective optics in the receivers are shown in Figure \ref{fig:strehls} for a representative observing wavelength $\lambda = 1.2$~mm.  
We note that the full 2$^\circ$ FoV is recovered at $\lambda\geq 2$~mm. Details and further optimizations will be discussed in a design release publication at the completion of the AtLAST design study.

\begin{figure*}[htb]
  \centering
  \includegraphics[width=\textwidth]{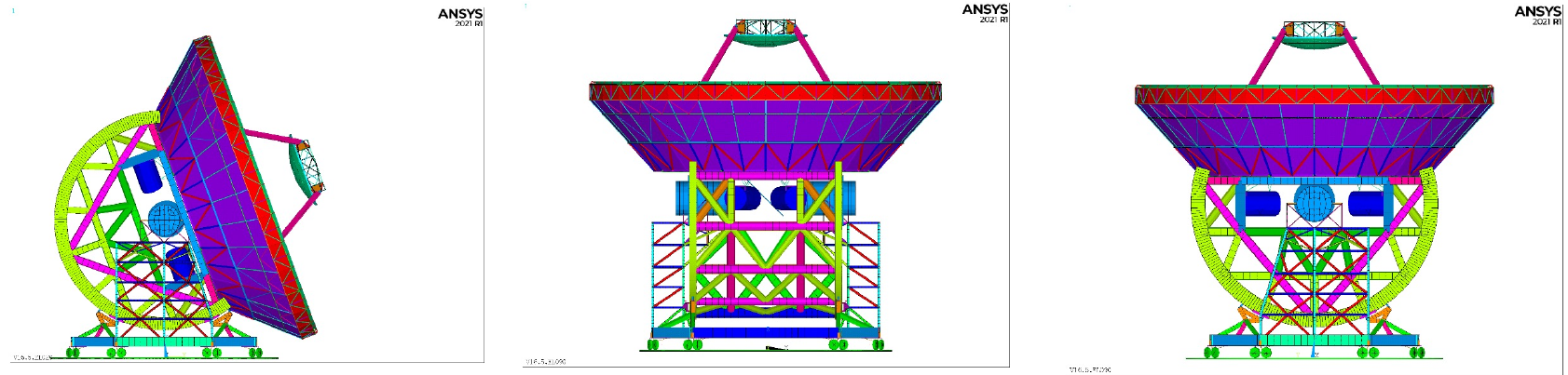}
  \caption{AtLAST finite element model, simulated in the Ansys Mechanical software package, in December 2022.}
  \label{fig:fe_model}
\end{figure*}

While the structural design initially proposed for AtLAST was based on a concept similar to that of the LMT/GTM \cite{2000SPIE.4015..155K}, the design quickly evolved to a rocking chair concept, which was conceived in 1988  (see \cite{2018ASSL..447.....B, 1989SPIE.1114..320K}). 
The rocking chair concept is now being realized in the ongoing construction of the 39-meter Extremely Large Telescope (ELT), a European Southern Observatory facility that will deliver an unprecedented collecting area in the optical and near-infrared wavebands, and is planned for the Giant Magellan Telescope and Thirty Meter Telescope.
The current baseline structural design for AtLAST is shown in Figure~\ref{fig:fe_model}.

\section{Future Steps}

With the baseline structural design now complete, the immediate next steps are to refine it, incorporating wind and weather data collected on the site in Chile as part of the EU-funded project, and to engineer the closed-loop solution for active optics that meet AtLAST's surface accuracy requirements \cite{2020SPIE11445E..2FK}, keeping the telescope in focus in a fast control loop \cite{2014SPIE.9145E..03K, 2020SPIE11445E..1NT}. By the end of 2023, we will:
\vspace{-8pt}
\begin{itemize}[topsep=0pt,itemsep=0pt,parsep=1pt,leftmargin=*]
    \item finalize and document the design description;
    \item produce engineering analysis reports on the optical, stress, mechanical, thermal, system, and electrical engineering, including end-to-end simulations;
    \item document potential procurement pathways for all major subsystems;
    \item generate a plan for manufacturing, factory and on-site assembly, transportation, testing and verification, and quality assurance;
    \item deliver a full computer aided design (CAD) model for AtLAST.
\end{itemize} 
\vspace{-8pt}
Following completion of these steps, we expect to deliver a complete AtLAST telescope design publication in 2024.

A large project like this necessitates strong international collaboration to thrive, and long-term financial and environmental sustainability of research infrastructures demand that efforts are merged, where possible, so that a larger community can benefit from a transformational facility at the same time minimizing its impact on the planet. Therefore, discussions are now underway to join the AtLAST and the 50-m Large Submillimeter Telescope (LST; \cite{2016SPIE.9906E..26K}) efforts, which closely parallel each other, after the AtLAST EU design is completed. Notably, AtLAST expands beyond the original goals of the LST in \cite{2016SPIE.9906E..26K} by designing for a $4\times$ larger FoV, a larger instrument cabin, and the ability to reach higher frequencies (1~THz vs the upper frequency of 420~GHz originally envisioned for LST).  The demand for these capabilities are not one-sided, as the LST community has also grown and evolved, generating compelling science cases that would benefit greatly from these high frequency capabilities. We are thus optimistic for the joint future of AtLAST/LST.

\section*{Acknowledgments}

We are grateful for the guidance and inspiration from Richard Hills throughout the early phases of the telescope design.\footnote{\url{https://atlast-telescope.org/news-and-events/news/2022/The-AtLAST-team-says-goodbye-to-Richard-Hills.html}}
This project has received funding from the European Union’s Horizon 2020 research and innovation program under grant agreement No.\ 951815 (AtLAST).

\bibliography{report} 

\begin{thebibliography}{10}

\bibitem{2022SPIE12190E..07R}
Joanna {Ramasawmy}, Pamela~D. {Klaassen}, Claudia {Cicone}, Tony~K.
  {Mroczkowski}, Chian-Chou {Chen}, et~al., ``{The Atacama Large Aperture
  Submillimetre Telescope: key science drivers},'' in [{\em Millimeter,
  Submillimeter, and Far-Infrared Detectors and Instrumentation for Astronomy
  XI}{\nolinebreak\hspace{0.1em}]},  Jonas {Zmuidzinas} and Jian-Rong {Gao},
  eds., {\em Society of Photo-Optical Instrumentation Engineers (SPIE)
  Conference Series} {\bf 12190},  1219007 (Aug. 2022).

\bibitem{2022arXiv221100195C}
John {Carpenter}, Crystal~L. {Brogan}, Daisuke {Iono}, and Tony {Mroczkowski},
  ``{The ALMA2030 Wideband Sensitivity Upgrade},'' {\em ALMA Memo 621} ,
  \url{https://library.nrao.edu/public/memos/alma/main/memo621.pdf} (Oct.
  2022).

\bibitem{2018SPIE10700E..41P}
Stephen~C. {Parshley}, Michael {Niemack}, Richard {Hills}, Simon~R. {Dicker},
  Rolando {D{\"u}nner}, et~al., ``{The optical design of the six-meter
  CCAT-prime and Simons Observatory telescopes},'' in [{\em Ground-based and
  Airborne Telescopes VII}{\nolinebreak\hspace{0.1em}]},  Heather~K. {Marshall}
  and Jason {Spyromilio}, eds., {\em Society of Photo-Optical Instrumentation
  Engineers (SPIE) Conference Series} {\bf 10700},  1070041 (July 2018).

\bibitem{2019arXiv190703479G}
Christopher {Groppi}, Andrey {Baryshev}, Urs {Graf}, Martina {Wiedner}, Pamela
  {Klaassen}, et~al., ``{First Generation Heterodyne Instrumentation Concepts
  for the Atacama Large Aperture Submillimeter Telescope},'' {\em arXiv
  e-prints} ,  arXiv:1907.03479 (July 2019).

\bibitem{2020SPIE11445E..2FK}
Pamela~D. {Klaassen}, Tony~K. {Mroczkowski}, Claudia {Cicone}, Evanthia
  {Hatziminaoglou}, Sabrina {Sartori}, et~al., ``{The Atacama Large Aperture
  Submillimeter Telescope (AtLAST)},'' in [{\em Society of Photo-Optical
  Instrumentation Engineers (SPIE) Conference
  Series}{\nolinebreak\hspace{0.1em}]},  {\em Society of Photo-Optical
  Instrumentation Engineers (SPIE) Conference Series} {\bf 11445},  114452F
  (Dec. 2020).

\bibitem{Gallardo2022}
Patricio~A. {Gallardo}, Bradford {Benson}, John {Carlstrom}, Simon~R. {Dicker},
  Nick {Emerson}, et~al., ``{Optical design concept of the CMB-S4
  large-aperture telescopes and cameras},'' in [{\em Millimeter, Submillimeter,
  and Far-Infrared Detectors and Instrumentation for Astronomy
  XI}{\nolinebreak\hspace{0.1em}]},  Jonas {Zmuidzinas} and Jian-Rong {Gao},
  eds., {\em Society of Photo-Optical Instrumentation Engineers (SPIE)
  Conference Series} {\bf 12190},  121900C (Aug. 2022).

\bibitem{AtLAST_memo_1}
R.~{Hills}, ``{AtLAST – options for the basic layout of the telescope},''
  {\em AtLAST Memo 1} ,
  {https://atlast--telescope.org/documents/memo--series/memo--public/basic--layout--options--v2.pdf}
  (May 2021).

\bibitem{2000SPIE.4015..155K}
Hans~J. {Kaercher} and Jacob~W. {Baars}, ``{Design of the Large Millimeter
  Telescope/Gran Telescopio Millimetrico (LMT/GTM)},'' in [{\em Radio
  Telescopes}{\nolinebreak\hspace{0.1em}]},  Harvey~R. {Butcher}, ed., {\em
  Society of Photo-Optical Instrumentation Engineers (SPIE) Conference Series}
  {\bf 4015},  155--168 (July 2000).

\bibitem{2018ASSL..447.....B}
Jacob W.~M. {Baars} and Hans~J. {K{\"a}rcher},  [{\em {Radio Telescope
  Reflectors: Historical Development of Design and
  Construction}}{\nolinebreak\hspace{0.1em}]}, vol.~447 (2018).

\bibitem{1989SPIE.1114..320K}
H.~J. {K{\"a}rcher} and H.~{Nicklas}, ``{Active structural control of very
  large telescopes.},'' in [{\em Active telescope
  systems}{\nolinebreak\hspace{0.1em}]},  Francois~J. {Roddier}, ed., {\em
  Society of Photo-Optical Instrumentation Engineers (SPIE) Conference Series}
  {\bf 1114},  320--327 (Sept. 1989).

\bibitem{2014SPIE.9145E..03K}
Hans~J. {K{\"a}rcher} and Jacob W.~M. {Baars}, ``{Ideas for future large single
  dish radio telescopes},'' in [{\em Ground-based and Airborne Telescopes
  V}{\nolinebreak\hspace{0.1em}]},  Larry~M. {Stepp}, Roberto {Gilmozzi}, and
  Helen~J. {Hall}, eds., {\em Society of Photo-Optical Instrumentation
  Engineers (SPIE) Conference Series} {\bf 9145},  914503 (July 2014).

\bibitem{2020SPIE11445E..1NT}
Yoichi {Tamura}, Ryohei {Kawabe}, Yuhei {Fukasaku}, Kimihiro {Kimura},
  Tetsutaro {Ueda}, et~al., ``{Wavefront sensor for
  millimeter/submillimeter-wave adaptive optics based on aperture-plane
  interferometry},'' in [{\em Society of Photo-Optical Instrumentation
  Engineers (SPIE) Conference Series}{\nolinebreak\hspace{0.1em}]},  {\em
  Society of Photo-Optical Instrumentation Engineers (SPIE) Conference Series}
  {\bf 11445},  114451N (Dec. 2020).

\bibitem{2016SPIE.9906E..26K}
Ryohei {Kawabe}, Kotaro {Kohno}, Yoichi {Tamura}, Tatsuya {Takekoshi}, Tai
  {Oshima}, et~al., ``{New 50-m-class single-dish telescope: Large
  Submillimeter Telescope (LST)},'' in [{\em Ground-based and Airborne
  Telescopes VI}{\nolinebreak\hspace{0.1em}]},  Helen~J. {Hall}, Roberto
  {Gilmozzi}, and Heather~K. {Marshall}, eds., {\em Society of Photo-Optical
  Instrumentation Engineers (SPIE) Conference Series} {\bf 9906},  990626 (Aug.
  2016).

\end{thebibliography}
\bibliographystyle{spiebib_5} 

\end{document}